\documentclass[conference]{IEEEtran}
\IEEEoverridecommandlockouts
\usepackage{cite}
\usepackage{amsmath,amssymb,amsfonts}
\usepackage{algorithmic}
\usepackage{graphicx}
\usepackage{textcomp}
\usepackage{xcolor}
\usepackage[colorlinks=true]{hyperref}
\usepackage{float}
\def\BibTeX{{\rm B\kern-.05em{\sc i\kern-.025em b}\kern-.08em
    T\kern-.1667em\lower.7ex\hbox{E}\kern-.125emX}}
    
\begin{document}

\IEEEoverridecommandlockouts            % <-- lets the \thanks in \title work
\title{A New DAPO Algorithm for Stock Trading%
  \thanks{Accepted at the 2025 IEEE 11th International Conference on Intelligent Data and Security (IDS). 
  Ranked 2\textsuperscript{nd} place in the FinRL Contest 2025 (Task 1).}}

\author{%
\IEEEauthorblockN{Ruijian Zha$^{\dagger}$}
\IEEEauthorblockA{\textit{Department of Computer Science, Columbia University}\\
New York, NY, USA\\
rz2689@columbia.edu}
\and
\IEEEauthorblockN{Bojun Liu$^{\dagger}$}
\IEEEauthorblockA{\textit{Department of Computer Science, Columbia University}\\
New York, NY, USA\\
bl3092@columbia.edu}
\thanks{$^{\dagger}$These authors contributed equally to this work.}
}

\maketitle

\begin{abstract}
Recent advances in reinforcement learning, like Dynamic sAmpling Policy Optimization (DAPO) algorithm, have demonstrated strong performance when combined with Large Language Models (LLMs). Motivated by this, we investigate whether similar benefits can be achieved in financial trading. We develop a novel trading agent that integrates an \textit{improved} Group Relative Policy Optimization (GRPO) algorithm, enhanced with insights from DAPO, and incorporates LLM-based signals of risk and sentiment from financial news. We evaluate our method on the NASDAQ-100 index using the FNSPID dataset. Our improved DAPO algorithm achieves a cumulative return of 230.49 \% and an Information Ratio of 0.37, outperforming the CPPO-DeepSeek baseline. Additionally, our approach reduces training time from approximately 8 hours to 2.5 hours over 100 epochs, while significantly decreasing RAM usage. Our approach addresses limitations of existing RL-LLM frameworks, offering a scalable solution for building financial trading agents. Code is available at: \url{https://github.com/Ruijian-Zha/FinRL-DAPO-SR/}
\end{abstract}


\section{Introduction}
Financial Reinforcement learning (FinRL) has gained traction in algorithmic trading for automating portfolio decisions under uncertainty \cite{b1,b2}. However, challenges like large drawdowns and limited interpretability persist. Extensions of PPO, e.g., CPPO \cite{b3} and FinRL-DeepSeek \cite{b5}, incorporate tail-risk constraints and textual signals, yet often require lengthy training (\(>7\) hours) and significant memory overhead.

The FinRL Contest 2025 \cite{b10} aims to advance the application of reinforcement learning in finance by evaluating intelligent trading strategies. \textit{Task 1: FinRL-DeepSeek} \cite{b10} explores the integration of FinRL with large language models (LLMs), using DeepSeek’s open-source models \cite{b4} to generate sentiment and risk signals from financial news. This task encourages innovation in combining quantitative modeling with natural language understanding to improve trading performance.

We made the following three contributions:
\begin{itemize}
    \item Adopting \textbf{Group Relative Policy Optimization (GRPO)}, a critic-free RL approach that reduces memory usage by sampling multiple actions from each state and normalizing the rewards within that group.
    \item Leveraging \textbf{Decoupled Clipping and Dynamic sAmpling Policy Optimization (DAPO)} \cite{b7}, originally proposed for LLM preference tuning, to avoid issues like entropy collapse and to enable dynamic sampling of meaningful state-action pairs.
    \item Introducing an \textbf{adjustable reward formula} that allows more nuanced control. This enables us to emphasize sentiment or risk as desired.
\end{itemize}
Our tests on the NASDAQ-100 index show that our new DAPO algorithm surpasses the best performance of CPPO-DeepSeek \cite{b4} while significantly reduces computational requirements.

\section{Related Work}\label{sec:related}
\subsection{Reinforcement Learning for Trading}
PPO \cite{b1} is widely applied to stock trading scenarios \cite{b2}, but it can be vulnerable to large drawdowns. CPPO \cite{b3} addresses tail risk using Conditional Value at Risk (CVaR) constraints, and FinRL-DeepSeek \cite{b5} extends CPPO by incorporating LLM-generated sentiment and risk signals from financial news data. While effective, these methods are typically resource-intensive due to high-dimensional state spaces, value functions, and extensive hyperparameter tuning.

\subsection{GRPO and DAPO}
Group Relative Policy Optimization (GRPO) eliminates the need for a value function by calculating group-level advantages, promising lower memory consumption and stable updates.  
\textbf{DAPO} \cite{b7} refines GRPO for large-scale language models, introducing:
\begin{itemize}
    \item \emph{Decoupled clipping}: Replaces symmetric clipping with an asymmetric range defined by parameters, $\epsilon_{\text{low}}$ and $\epsilon_{\text{high}}$. This allows flexible policy updates, enhancing exploration in high-reward scenarios while constraining risk.
    \item \emph{Dynamic sampling}: Samples with uniform risk-adjusted rewards are filtered out, ensuring that the model focuses on variations where learning signals are richer, thus promoting faster and more stable convergence.

\end{itemize}
We adapt these insights to a FinRL setting, focusing on daily aggregated sentiment and risk signals from LLMs.

\section{Methodology}\label{sec:method}
\subsection{Stock Trading Environment}
We follow the standard FinRL Contest setting for the definition of states, actions, and rewards as outlined in \cite{b10}. The state represents the current status of the trading environment, including the available cash, stock prices, number of shares held, and various technical indicators, including LLM sentiment and risk values. The action refers to the decisions made by the trading agent, which includes buying, selling or holding stocks. The reward is the feedback signal received after taking an action, calculated as the change in total asset value.

\subsection{GRPO with Exponentiated Sentiment-Risk Reward}
\textbf{Group Relative Policy Optimization (GRPO).}
For each state \(s_t\), the policy \(\pi_\theta\) samples a group of \(n\) potential actions \(\{a_{t,1}, a_{t,2}, \dots, a_{t,n}\}\). The group advantage for action \(a_{t,i}\) is then defined as:
\begin{equation}
A^G(s_t, a_{t,i}) = \frac{r_{t,i} - \mu_t}{\sigma_t + \epsilon}\end{equation}
where $r_{t,i}$ denotes the reward for each candidate action, $\epsilon$ is a small constant to avoid division by zero, and 
\begin{equation}
\quad
\mu_t = \frac{1}{n} \sum_{j=1}^{n} r_{t,j}, \quad
\sigma_t = \sqrt{\frac{1}{n}\sum_{j=1}^{n} (r_{t,j} - \mu_t)^2}.
\end{equation}
By relying on these group-based comparisons, GRPO obviates the need for a separate critic network, reducing computational overhead and potentially stabilizing updates.

\textbf{Sentiment-Risk Adjusted Rewards.}
Our baseline reward \(r_{t,i}\) for action \(a_{t,i}\) is the portfolio return over one trading period. Let \(S_{t,i}\) and \(R_{t,i}\) be the aggregated sentiment and risk scores for each stock \(i\) at time \(t\).  We then incorporate sentiment and risk via:
\begin{equation}
r'_{t,i} = r_{t,i} \cdot \frac{\bigl(S_{t,i}\bigr)^{\alpha}}{\bigl(R_{t,i}\bigr)^{\beta}+1e-8}
\end{equation}
where
\[
S_{t,i} = \sum_{j=1}^{m} w_{t,i,j} \, f(S_{f,j}), \quad
R_{t,i} = \sum_{j=1}^{m} w_{t,i,j} \, f(R_{f,j}),
\]
and weight \(w_{t,i,j}\) reflects the proportion of stock \(j\) held by action \(a_{t,i}\). Meanwhile, \(f\) map discrete LLM-generated sentiment and risk scores (\(1\ldots5\)) to continuous factors (\ \(0.99, 0.995, 1.0, 1.005, 1.01\)), amplifying rewards under favorable sentiment and low risk. The exponents \(\alpha,\beta \ge 0\) control the relative influence of sentiment vs.\ risk; for instance, \(\alpha > \beta\) gives heavier weight to sentiment, while \(\alpha = \beta = 0\) effectively disables sentiment-risk adjustments.

\subsection{GRPO-Inspired DAPO}
\textbf{Policy Optimization with Decoupled Clipping.}
While GRPO offers a critic-free and memory-efficient alternative to traditional RL methods, it still suffers from limited exploration and learning inefficiencies in flat reward landscapes. To address these issues, we replace the uniform symmetric clipping parameter $\epsilon$ in GRPO with two asymmetric thresholds: $\epsilon_{low}$ and $\epsilon_{high}$. The GRPO loss is updated as:
\begin{flalign}
&\mathcal{L}^{GRPO}(\theta) \nonumber \\ 
&= \mathbb{E} \Big[ 
\min \big( r_t(\theta)\, A^G_t,\;
\mathrm{clip}\big(r_t(\theta), 1-\epsilon_{\text{low}},\, 1+\epsilon_{\text{high}}\big) A^G_t \big) 
\Big],
\end{flalign}
where $r_t(\theta) = \frac{\pi_\theta(a_t \mid s_t)}{\pi_{\theta_{\mathrm{old}}}(a_t \mid s_t)}$. By enabling asymmetric clipping, the policy has greater flexibility to reinforce advantageous actions, while large negative updates remain constrained for stability.

\textbf{Dynamic Sampling.}
To improve sample efficiency, we filter out states with all the same rewards. The policy parameters $\theta$ are updated via stochastic gradient descent:
\[
\theta \leftarrow \theta + \alpha \nabla_{\theta} \mathcal{L}^{GRPO}(\theta).
\]
By filtering out uninformative samples and employing decoupled clipping, our GRPO-inspired DAPO algorithm balances exploration and stability in large-scale financial trading.

\section{Preliminary Results and Discussion}\label{sec:results}

\subsection{Dataset}
We use the FNSPID dataset \cite{b8}, which includes 15.7 million time-aligned financial news records from 1999–2023. Since our primary objective is not to derive sentiment and risk signals, we adopt the pre-extracted LLM-based metrics from FinRL-DeepSeek \cite{b5}, avoiding the overhead of retraining or fine-tuning large language models.

\subsection{Impacts of Risk and Sentiment Signals}
We first assess the contribution of risk and sentiment signals by comparing four configurations: (1) using only risk signals (\(\alpha = 0, \beta = 1\)), (2) only sentiment signals (\(\alpha = 1, \beta = 0\)), (3) balanced risk and sentiment (\(\alpha = 1, \beta = 1\)), with (4) the NASDAQ-100 index as a baseline. Fig~1 shows that the balanced configuration achieves a higher cumulative returns and improved risk-adjusted metrics compared to using either signal alone. Incorporating both signals allows the agent to favor actions with high sentiment (indicating market optimism) while avoiding high-risk scenarios, leading to more robust trading decisions.

\begin{figure}[H]
    \centering
    \includegraphics[width=\linewidth]{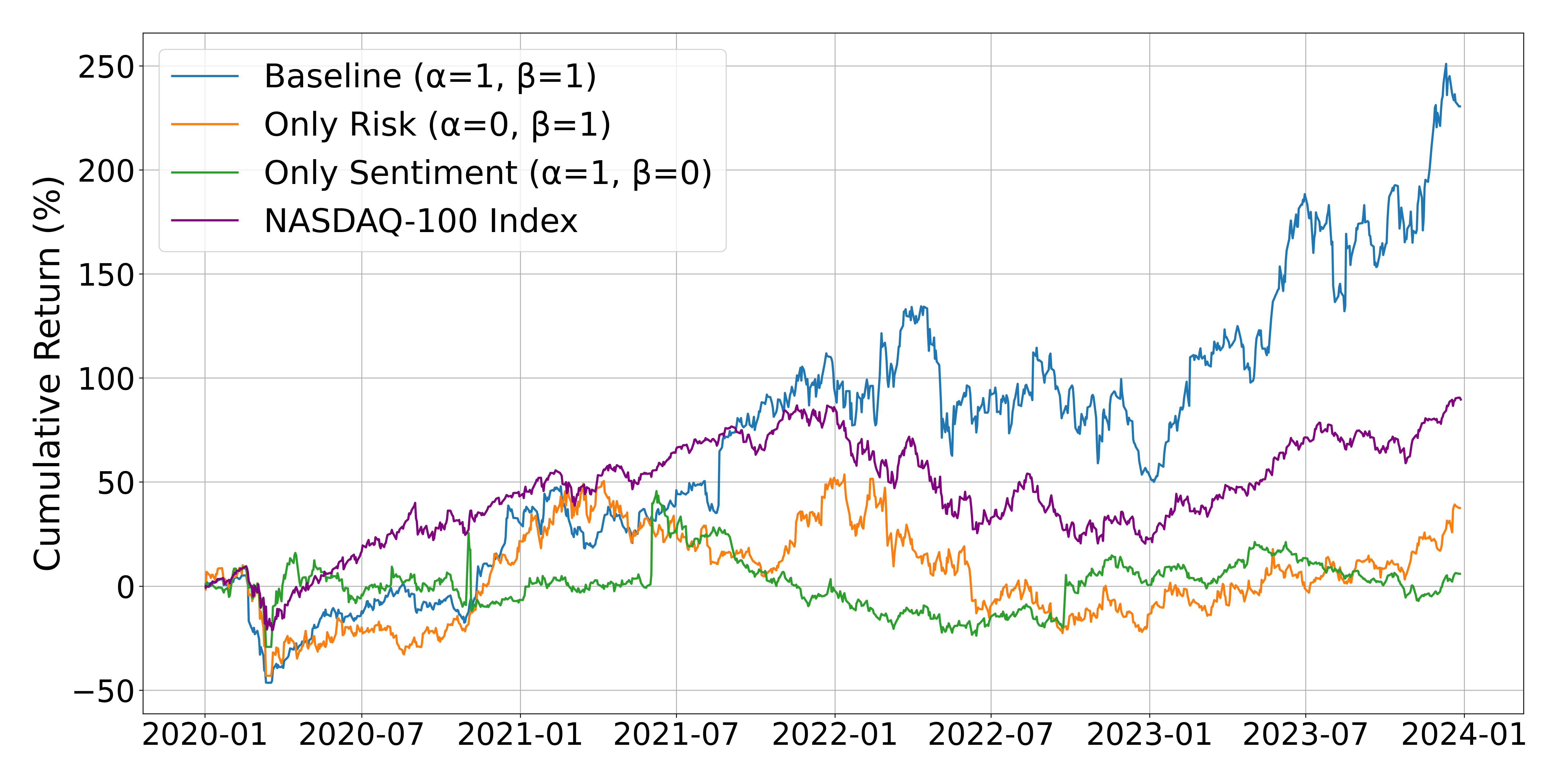}
    \caption{Comparison of the Baseline, Risk‐Only, and Sentiment‐Only reward settings vs. the NASDAQ‐100, showing 2020–2023 cumulative returns after 100 training epochs on six years of data.}
    \label{fig:enter-label}
\end{figure}

\subsection{Effects of Varying \(\alpha\) and \(\beta\)}
Next, we evaluate how different exponents for sentiment (\(\alpha\)) and risk (\(\beta\)) influence the reward shaping. We compare five settings: (1) Risk-Heavy (\(\alpha = 1, \beta = 3\)), (2) Balanced (\(\alpha = 2, \beta = 2\)), (3) Sentiment Emphasis (\(\alpha = 3, \beta = 1\)), (4) Strong Sentiment (\(\alpha = 5, \beta = 1\)), and (5) the NASDAQ-100 index. Fig~2 illustrates that a sentiment-emphasized configuration \((\alpha = 3, \beta = 1)\) yields notably higher cumulative returns than either balanced or risk-focused settings, suggesting that strong sentiment signals can drive profitable trades.

\begin{figure}[H]
    \centering
    \includegraphics[width=1\linewidth]{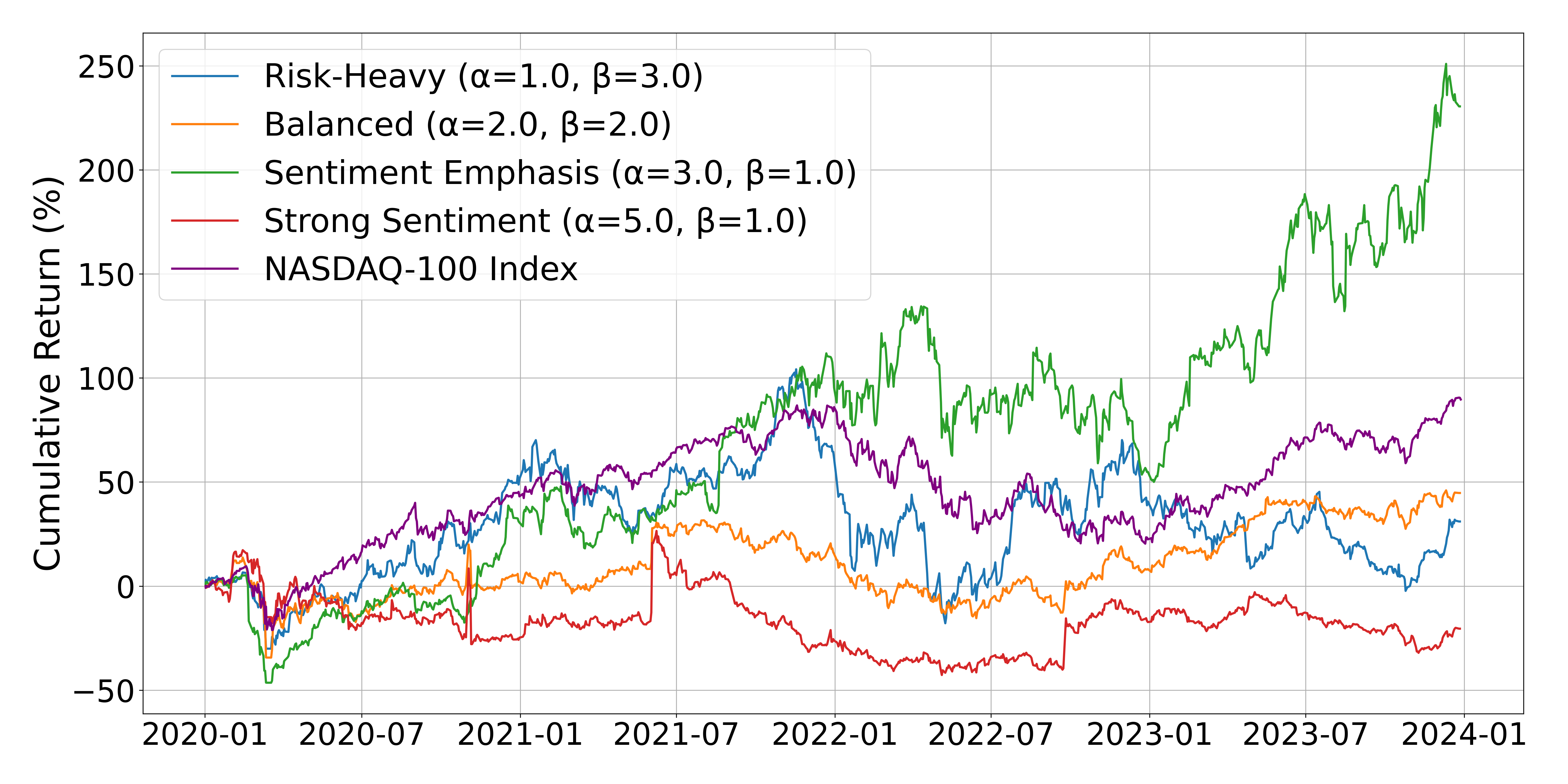}
    \caption{Comparison of four sentiment–risk weightings vs. the NASDAQ‐100, showing 2020–2023 cumulative returns after 100 training epochs on six years of data.}
    \label{fig:enter-label}
\end{figure}

Both balanced (\(\alpha=1, \beta=1\)) and Sentiment Emphasis (\(\alpha=3, \beta=1\)) achieve $\sim$ 230.49\% return, reflecting the reward function’s robustness within a reasonable \(\alpha, \beta\) range. Moderate \(\alpha\) values leverage sentiment effectively, while \(\beta=1\) ensures balanced risk consideration. In contrast, Risk-Heavy (\(\alpha=1, \beta=3\)) over-penalizes risk, and Strong Sentiment (\(\alpha=5, \beta=1\)) over-emphasizes sentiment, leading to suboptimal results. This indicates \(\alpha\) and \(\beta\) should remain balanced to maintain consistent performance across market conditions.

\begin{table}[htbp]
\begin{center}
\begin{tabular}{lcc}
\hline
\textbf{Metric} & \textbf{Our Model} & \textbf{CPPO-DeepSeek 10\%} \\
\hline
Cumulative Return          & \textbf{230.49\%}     & $\sim$215\% \\
Max Drawdown               & -49.11\%              & \textbf{$\sim$-35\%} \\
Rachev Ratio\textsuperscript{1}              & \textbf{1.12}         & 0.9818 \\
Information Ratio\textsuperscript{2}        & \textbf{0.37}         & 0.0078 \\
CVaR (5\%)                 & -5.64\%               & \textbf{-4.37\%} \\
Outperformance Frequency   & 50.0\%                & Not reported \\
\hline
\end{tabular}
\end{center}
{\textit{Table 1}. Comparison of key metrics between our method and CPPO-DeepSeek 10\%, which incorporates DeepSeek data.}
% \label{tab:final_comparison}
\end{table}

To comply with the contest requirement, we also evaluated our model on the 2019--2023 period. 
Table~2 summarizes the performance metrics over this timeframe:

\begin{table}[htbp]
\begin{center}
\begin{tabular}{lcc}
\hline
\textbf{Metric} & \textbf{Our Model (2019--2023)} \\
\hline
Cumulative Return          & 335.58\% \\
Max Drawdown               & -50.24\% \\
Rachev Ratio               & 1.09     \\
Information Ratio          & 0.30     \\
CVaR (5\%)                 & -5.50\%  \\
Outperformance Frequency   & 49.6\%   \\
\hline
\end{tabular}
\end{center}
{\textit{Table 2.} Backtesting results of our algorithm for the 2019--2023 period, meeting the contest’s guideline for evaluation.}
\label{tab:backtest_2019_2023}
\end{table}

\subsection{Improved Computation Efficiency}
Our DAPO algorithm significantly reduces computational requirements compared to the CPPO-DeepSeek \cite{b5}. Experiments were conducted on a system with an Intel(R) Xeon(R) CPU @ 2.20GHz, 12 logical cores, 6 physical cores.

\begin{table}[H]
\begin{center}
\begin{tabular}{lcc}
\hline
\textbf{Metric} & \textbf{Our Model} & \textbf{CPPO-DeepSeek 10\%} \\
\hline
RAM Usage (GB)             & \textbf{15}       & 120 \\
Training Time (100 Epochs) & \textbf{2.5 hours} & $\sim$7-8 hours \\
\hline
\end{tabular}
\end{center}
{\textit{Table 3}. Comparison of memory and training time (100 epochs, 20k steps each).}
\end{table}

\section{Conclusion and Future Work}\label{sec:conclusion}
We presented an improved DAPO algorithm for algorithmic trading that integrates exponent-weighted sentiment-risk scores. This approach allows flexible tuning of the reward to emphasize sentiment or risk. Results on the NASDAQ-100 index demonstrate potential for superior returns vs. prior SOTA, with faster training and much less memory usage. 

Future directions include:
\begin{enumerate}
    \item \textbf{Intraday Timescales}: Extending the approach to shorter intervals for rapid event-driven trading.
    \item \textbf{Enhanced Prompts}: Using more detailed LLM prompts for sector-specific or event-specific signals.
    \item \textbf{Adaptive \(\alpha,\beta\)}: Automatically adjusting exponents based on market regimes.
\end{enumerate}

\footnotetext[1]{The Rachev Ratio measures tail-risk-adjusted performance by comparing expected tail gains to expected tail losses.}
\footnotetext[2]{Information Ratio evaluates the consistency of excess returns relative to a benchmark.}

\begin{thebibliography}{00}
\bibitem{b1} J. Schulman, F. Wolski, P. Dhariwal, A. Radford, and O. Klimov, 
``Proximal policy optimization algorithms,'' \emph{arXiv preprint arXiv:1707.06347}, 2017.

\bibitem{b2} X.-Y. Liu, H. Yang, Q. Chen, \emph{et al.}, 
``FinRL: A deep reinforcement learning library for automated stock trading in quantitative finance,'' 
\emph{arXiv preprint arXiv:2011.09607}, 2022.

\bibitem{b3} C. Y. Ying, X. Zhou, H. Su, \emph{et al.},
``Towards safe reinforcement learning via constraining conditional value-at-risk,''
in \emph{IJCAI}, 2022, pp.~3673--3680.

\bibitem{b4} DeepSeek-AI \emph{et al.},
``DeepSeek-V3 Technical Report,''
\emph{arXiv preprint arXiv:2412.19437}, 2024.

\bibitem{b5} M. Benhenda, 
``FinRL-DeepSeek: LLM-infused risk-sensitive reinforcement learning for trading agents,''
\emph{arXiv preprint arXiv:2502.07393}, 2025.

\bibitem{b7} Q. Yu, Z. Zhang, R. Zhu, \emph{et al.},
``DAPO: An open-source LLM reinforcement learning system at scale,''
\emph{arXiv preprint arXiv:2503.14476}, 2025.

\bibitem{b8} Z. Dong, X. Fan, Z. Peng, \emph{et al.}, 
``Fnspid: A comprehensive financial news dataset in time series,''
\emph{arXiv preprint arXiv:2402.06698}, 2024.

\bibitem{b9} C. Y. Ying, X. Zhou, H. Su, D. Yan, N. Chen, and J. Zhu, 
``Towards safe reinforcement learning via constraining conditional value-at-risk,''
in \emph{Proceedings of the Thirty-First International Joint Conference on Artificial Intelligence (IJCAI-22)}, 
pp. 3673–3680, 2022, 
doi: 10.24963/ijcai.2022/510.

\bibitem{b10} K. Wang, K. Xiao, and X.Y. Liu, 
``Parallel Market Environments for FinRL Contests,''
\emph{arXiv preprint arXiv:2504.02281}, 2025.

\bibitem{b11} X.-Y. Liu, Z. Xia, H. Yang, J. Gao, D. Zha, M. Zhu, C. D. Wang, Z. Wang, and J. Guo,``Dynamic datasets and market environments for financial reinforcement learning, '' \emph{Machine Learning - Nature}, 2024.

\bibitem{b11} H. H. Yang, X.-Y. Liu, H. Tong, Y. Zhang, J. Wang, and L. Deng, ``Deep reinforcement learning for automated stock trading: an ensemble strategy,'' \emph{Proceedings of the 30th ACM International Conference on Information \& Knowledge Management (CIKM)}, pp. 2245--2252, 2020.


\end{thebibliography}
\end{document}